\documentclass[aps,prb,notitlepage,twocolumn,10pt,superscriptaddress]{revtex4-2}
\usepackage{graphicx}
\usepackage{amsmath}
\usepackage{amssymb}
\usepackage{epstopdf}
\usepackage[colorlinks=true,linkcolor=blue,citecolor=blue]{hyperref}
\usepackage{xcolor}
\usepackage{comment}
\usepackage{tabu,hhline}

\begin{abstract}
Multiferroic materials, characterized by the occurrence of two or more ferroic properties, hold potential in future technological applications and also exhibit intriguing phenomena caused by the interplay of multiple orders. One such example is the formation of spin cycloid structures within multiferroic materials, which we investigate in this work by focusing on their magnon excitations and transport based on a general multiferroic Hamiltonian with an antiferromagnetic order. More specifically, we identify the ground state and explore the dynamics of magnon modes, revealing distinct in-plane and out-of-plane modes with anisotropic dispersion relations.
The magnon modes include a massless excitation, known as the Goldstone boson, originating from the spontaneous breaking of the translational symmetry by the formation of the cycloid structures. By employing the Boltzmann transport formalism, the magnonic thermal conductivity with spin cycloids and low-temperature anisotropic behaviors is discussed. This work provides pathways to envision the spin-textured multiferroics, which may serve as a fertile ground to look for novel thermal and spin transport with the rich interplay of quasiparticles such as magnons and phonons.
\end{abstract}

\begin{document}

\author{Hyeon Woo Park}
\affiliation{Department of Physics, Korea Advanced Institute of Science and Technology, Daejeon 34141, Republic of Korea}

\author{Shu Zhang}
\affiliation{Institute for Theoretical Solid State Physics, IFW Dresden, Helmholtzstrasse 20, 01069 Dresden, Germany}

    \author{Peter Meisenheimer}
\affiliation{Department of Materials Science and Engineering, University of California, Berkeley, CA 94720, USA}

\author{Maya Ramesh}
\affiliation{Department of Materials Science and Engineering, Cornell University, Ithaca, NY 14853, USA}

\author{Sajid Husain}
\affiliation{Department of Materials Science and Engineering, University of California, Berkeley, CA 94720, USA}
\affiliation{Materials Science Division, Lawrence Berkeley National Laboratory, Berkeley, CA 94720, USA}

\author{Isaac Harris}
\affiliation{Department of Physics, University of California, Berkeley, CA 94720, USA}

\author{Jorge Íñiguez-González}
\affiliation{Materials Research and Technology Department, Luxembourg Institute of Science and Technology (LIST), Esch-sur-Alzette Luxembourg}
\affiliation{Department of Physics and Materials Science, University of Luxembourg, Esch-sur-Alzette, Belvaux 1511, Luxembourg}

\author{Zhi Yao}
\affiliation{Applied Mathematics and Computational Research Division, Lawrence Berkeley National Laboratory, Berkeley, CA 94720, USA}

\author{Ramamoorthy Ramesh}
\affiliation{Department of Materials Science and Engineering, University of California, Berkeley, CA 94720, USA}
\affiliation{Materials Science Division, Lawrence Berkeley National Laboratory, Berkeley, CA 94720, USA}
\affiliation{Department of Physics, University of California, Berkeley, CA 94720, USA}
\affiliation{Departments of Physics and Astronomy and Materials Science and Nano Engineering and Rice Advanced Materials Institute, Rice University, Houston, TX 77005, USA}

\author{Se Kwon Kim}
\affiliation{Department of Physics, Korea Advanced Institute of Science and Technology, Daejeon 34141, Republic of Korea}

\title{Magnon thermal conductivity in multiferroics with spin cycloids}
\date{\today}
\maketitle

\section{Introduction}
Multiferroics refer to substances that exhibit more than one order parameter, the study of which has formed a rapidly advancing field within materials science and spintronics~\cite{1982_SPU,2006_nature}.
The combination of multiple order parameters confers novel functionalities upon these materials, fostering potential applications in diverse fields, such as memory devices and sensors~\cite{2008_nat_mat,2014_nat,2019_nat_mat}.
Currently, extensive research efforts are underway to explore multiferroics across various substances, prominently the bismuth ferrite BiFeO$_3$ (BFO) that harbors ferroelectricity and antiferromagnetism simultaneously at room temperature~\cite{PhysRevLett.109.067203, MOREAU19711315,PhysRevB.73.132101,doi:10.1126/science.1080615,2006,2014_park}. One reason why multiferroics are attracting significant attention is that they enable the control of magnetism through an electric field~\cite{PhysRevLett.100.227602,10.1038/nature23656, 10.1038/nmat1731, 10.1038/s41467-020-15501-8, 10.1038/s41467-024-47232-5}.
In the case of typical antiferromagnets, the magnetic moments of the sublattices cancel each other out, making it challenging to manipulate and detect the magnetic order parameters for device applications. However, in multiferroics, the antiferromagnetic order that can be modulated and probed by electrical means through magnetoelectric coupling associated with the two coexisting order parameters~\cite{doi.org/10.1038/nmat2899,PhysRevLett.129.087601,PhysRevLett.114.157203,PhysRevLett.134.016703}.

In contrast to conventional ferromagnetic or ferroelectric materials where ground states are often uniform, multiferroic materials tend to exhibit a distinct ground state~\cite{SOSNOWSKA1996384}. Among the intriguing phenomena observed within various multiferroics, including BFO, is the manifestation of spin cycloid structures, characterized by the helical modulation of magnetic orders caused by the inversion-symmetry breaking associated with the electric polarization~\cite{https://doi.org/10.1002/adma.202003711, 2006,2014_park,SOSNOWSKA1995167}. Understanding such spin cycloidal structures and their effects on other physical properties such as heat and spin transport is crucial to elucidate the mechanisms underlying various phenomena occurring in multiferroic materials and also to their practical use~\cite{PhysRevLett.129.087601,10.1038/s41563-024-01854-8,PhysRevLett.114.157203,10.1038/s41467-024-50180-9}.

In this work, we provide a theoretical framework of the magnon transport in magnetoelectric multiferroics. Based on a general Landau-Ginzburg formalism for multiferroics with antiferromagnetic orders, we obtain the ground states and determine the associated magnon band structure in the presence of an external magnetic field.
The magnon band structure manifests the anisotropy in the spin cycloid structure. Along the direction of the order parameter winding, the spin cycloid defines a spatial period that imprints on the magnon bands, which results in a gap opening at the magnetic Brillouin zone boundary. The thermal transport conducted by magnons also shows an anisotropy, offering a transport probe of the microscopic spin cycloid texture. In addition, the dependence of the magnonic thermal conductivity on the magnetic field reveals the field-induced change of the underlying spin-cycloid structure, providing a way to probe the elusive background spin texture.

The remaining sections of the paper are organized as follows.
In Sec.~\ref{section2}, the model to calculate the thermal conductivity in multiferroics is described.
In Sec.~\ref{section3}, we discussed the results of the ground state of the systems, and use these results to calculate the excitations and anisotropic thermal conductivity followed by the conclusion Sec.~\ref{section4}.

\section{Model} \label{section2}
In this section, we elucidate the model construction of multiferroic materials, utilizing the Landau-Ginzberg free-energy density, represented as~\cite{PhysRevB.50.2953, PhysRevB.77.012406, 2014_park, Tehranchi01121997}
\begin{align} \label{free_energy}
    f_{l} =& A (\nabla \mathbf{n})^2 + \frac{\mathbf{m}^2}{2\chi} - \alpha\mathbf{P} \cdot [\mathbf{n}(\nabla \cdot\mathbf{n}) + \mathbf{n}\times(\nabla \times \mathbf{n})] \nonumber \\
    &-2\beta \hat{\mathbf{z}}\cdot(\mathbf{m}\times\mathbf{n}) - K_u n_z^2 -\mathbf{h}\cdot \mathbf{m},
\end{align}
where $\mathbf{n}$ is the antiferromagnetic vector, which has unit length and $\mathbf{m}$ is magnetization vector.
Additionally, the free energy is associated with the ferroelectric polarization $\mathbf{P}$.
The first term accounts for the exchange interaction, while the second term represents the suppression of the magnetization associated with antiferromagnetism. The third and fourth terms embody the Lifshitz invariant and Dzyaloshinskii–Moriya interaction, respectively.
The origin of the third term is the interaction between the electric polarization associated with the underlying ferroelectric order and the electric polarization induced by the magnetic texture~\cite{2005_Katsura,PhysRevLett.96.067601}.
The fourth term arises when inversion symmetry is broken in the crystal structure.
Note that, in the case of BFO, oxygen octahedral tilts involve symmetry breakings that yield the occurrence of Dzyaloshinskii-Moriya interactions (DMI)~\cite{PhysRevLett.109.037207, Zvezdin_2012,FENG20101765,2014_park}.
 Nevertheless, since these octahedral tilts are known to be strongly coupled to the polarization~\cite{PhysRevB.106.165122}, it is usually sufficient to include $\mathbf{P}$ in a basic model.
Finally, the fifth and last term introduce single-ion easy-axis anisotropy and Zeeman coupling to an external magnetic field $\mathbf{h}$, respectively. The inspiration behind the formulation of the aforementioned free energy (\ref{free_energy}) stems from the Hamiltonian of multiferroic materials, such as BFO~\cite{PhysRevLett.109.037207,2014_park}.
In our work, it is assumed that $\mathbf{P}$ is aligned along the z-direction and has a fixed magnitude and that the Dzyaloshinskii–Moriya vector is in the $\hat{z}$-direction~\cite{2014_park}.

By employing the aforementioned free energy (\ref{free_energy}), we derive the Lagrangian density for the system:
\begin{align} \label{Lagrangian}
    \mathcal{L} =& \, s\mathbf{m}\cdot(\mathbf{n}\times \dot{\mathbf{n}}) - f_l \nonumber \\
    =& \, s\mathbf{m}\cdot(\mathbf{n}\times \dot{\mathbf{n}}) - A(\nabla \mathbf{n})^2 - \frac{\mathbf{m}^2}{2\chi} \nonumber \\
    &+ \alpha\mathbf{P} \cdot [\mathbf{n}(\nabla \cdot\mathbf{n}) + \mathbf{n}\times(\nabla \times \mathbf{n})] \nonumber \\
    &+ 2\beta \hat{\mathbf{z}}\cdot(\mathbf{m}\times\mathbf{n}) + K_u n_c^2 + \mathbf{h}\cdot\mathbf{m}.
\end{align}
In this expression, the first term represents the kinetic term of the magnetic system~\cite{sachdev, PhysRevB.93.104408}, while the subsequent terms denote the potential terms. Since the electric polarization and other structural distortions are assumed to be fixed, the dynamical term of $\mathbf{P}$ is neglected.
By employing the Euler-Lagrange equations and information regarding the ground state, we derive the dynamics of magnons from the Lagrangian (\ref{Lagrangian}).

\section{Result} \label{section3}
In this section, we present the findings of our investigation into the multiferroic model that we constructed. We begin by determining its ground state of given free energy, followed by an exploration of excitations within the system and an analysis of its thermal conductivity.

\subsection{Ground state}

\begin{figure}
    \centering
    \includegraphics[width=0.45\textwidth]{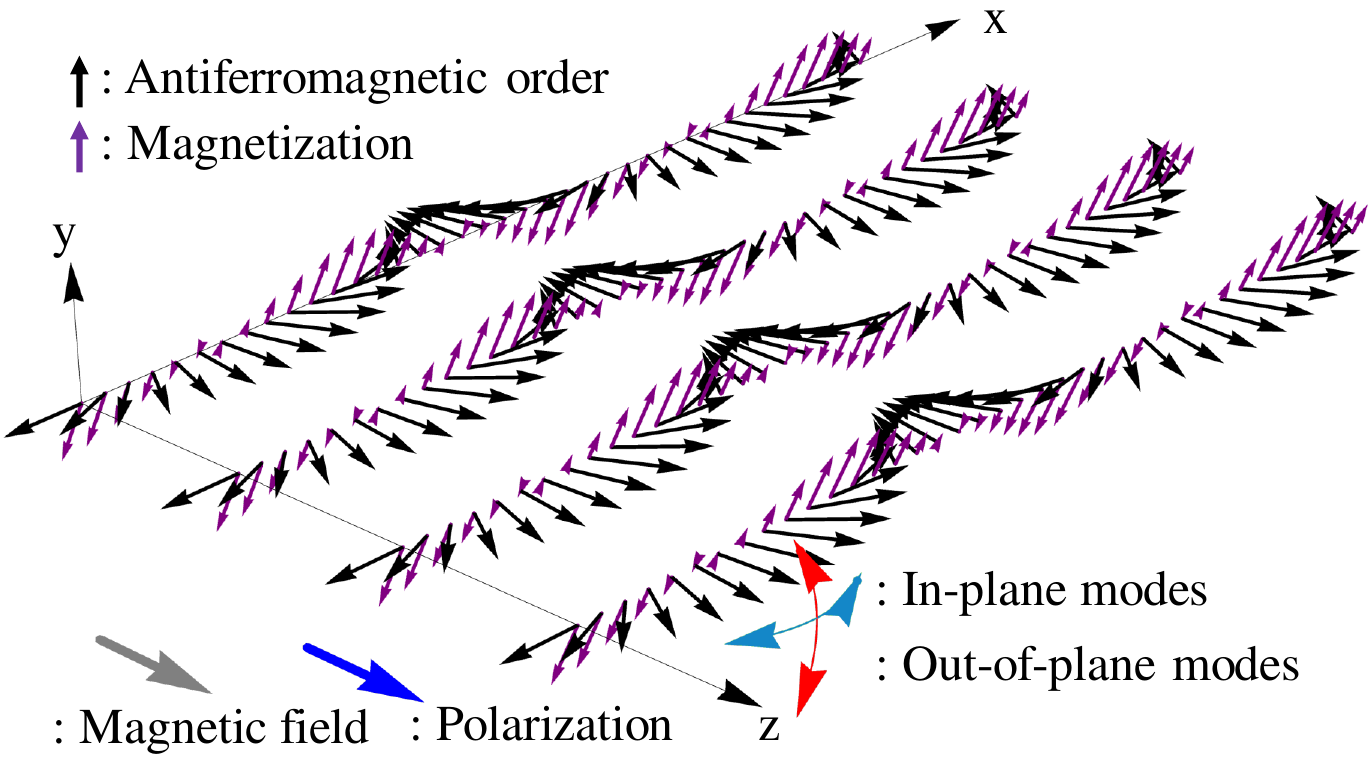}
    \caption{Ground state of the given free energy (\ref{free_energy}). Electric polarization and magnetic field are aligned in the z-direction. Purple arrows indicate the magnetization vector $\mathbf{m}$, while black arrows represent the unit antiferromagnetic vector $\mathbf{n}$.
    Magnetization vectors are depicted larger than its actual value and antiferromagnetic vectors are all placed on the $xz$-plane.
    The dynamics of the antiferromagnetic order parameter are depicted for the two magnon modes: the blue one represents the in-plane mode where the order parameter oscillates within the spin-cycloid plane and the red one represents the out-of-plane mode where the order parameter oscillates out of the spin-cycloid plane.}
    \label{fig:Ground state}
\end{figure}

To determine the ground state, we minimize the Landau-Ginzburg free energy under constraints $\left| \mathbf{n}\right| = 1$ and $\mathbf{n} \cdot \mathbf{m} = 0$.
The original constraint is $|\mathbf{n}|^2 + |\mathbf{m}|^2 = 1$, but under the assumption that the magnetization is very small, it simplifies to $\left| \mathbf{n}\right| = 1$.
For convenience, let us parameterize $\mathbf{m}$ and $\mathbf{n}$ as $\mathbf{n} = (\sin\theta\cos\phi,\sin\theta\sin\phi,\cos\theta)$ and $\mathbf{m} = m_\theta \hat{\mathbf{\theta}}+m_\phi \hat{\mathbf{\phi}}$, where  $\hat{\mathbf{\theta}} = (\cos\theta\cos\phi,\cos\theta\sin\phi,-\sin\theta)$ and $ \hat{\mathbf{\phi}} = (-\sin\phi,\cos\phi,0)$.
In the models we are letting $\mathbf{P} = (0, 0, P_z)$ and $\mathbf{h} = (0, 0, h)$. Then, the free-energy density in terms of the four dynamic variables $\theta, \phi, m_\theta, m_\phi$ is given by
\begin{align} \label{free energy 2}
    f_{l} =& A\left[(\nabla\theta)^2+\sin^2 \theta (\nabla\phi)^2\right] + \frac{m_\theta ^2 + m_\phi ^2}{2\chi} \nonumber \\
    &-\alpha P_z \{\cos\phi(\partial_x\theta)+\sin\phi(\partial_y\theta) \nonumber \\
    & \quad \quad \qquad -\sin\theta\cos\theta\left[\sin\phi(\partial_x\phi)-\cos\phi(\partial_y\phi)\right]\} \nonumber \\
    &+ 2\beta m_\phi\sin\theta - K_u \cos^2 \theta + h m_\theta \sin\theta.
\end{align}

Solving the Euler-Lagrange equation for given free energy (\ref{free energy 2}) about $\theta, \phi, m_\theta$ and $ m_\phi$ yields the following solutions:
\begin{align}
    & m_\phi = - 2\beta \chi \sin \theta, \quad m_\theta = -\chi h \sin\theta, \\
    & \nabla\phi = 0, \quad \sin\phi(\partial_x\theta)-\cos\phi(\partial_y\theta) = 0, \\
    & \phi = \text{constant} = \arctan\left(\frac{\partial_y\theta}{\partial_x\theta}\right), \\ 
    & 2A(\nabla^2\theta) - (K_u - 2\beta^2 \chi - h^2 \chi / 2) \sin 2\theta = 0 . \label{spin cycloid solution}
\end{align}
This solution creates a spin cycloid structure on the plane which is spanned by the z-axis and one specific direction in the $xy$-plane.
Here, we can see that the $K_u - 2\beta^2 \chi - h^2 \chi / 2$ term can be interpreted as the effective anisotropy of the system, leading us to define it as $K_{\text{eff}}$.
The spin cycloid structure can be analytically obtained using Jacobi elliptic functions :
\begin{align}
    \cos\theta &= \text{sn}\left(\sqrt{\frac{C}{A}}r,\frac{K_{\text{eff}}}{C}\right) , \\
    \sin\theta &= -\text{cn}\left(\sqrt{\frac{C}{A}}r,\frac{K_{\text{eff}}}{C}\right) ,
\end{align}
for the $K_{\text{eff}} > 0$ case, and 
\begin{align}
    \cos\theta &= \text{cn}\left(\sqrt{\frac{C}{A}}r,\frac{K_{\text{eff}}}{C}\right) , \\
    \sin\theta &= \text{sn}\left(\sqrt{\frac{C}{A}}r,\frac{K_{\text{eff}}}{C}\right) ,
\end{align}
for the $K_{\text{eff}} < 0$ case where C is the integration constant determined by minimizing the total free energy of the system. This is one of our main results: the analytical solution for spin cycloids of multiferroic materials [Eq.~(\ref{free_energy})] in the presence of an external magnetic field $h$. We note here that our solution generalizes the known results in the absence of a magnetic field which can be found in Refs.~\cite{2014_park, Tehranchi01121997, SOSNOWSKA1995167}.

Note that the DMI $\propto \beta$ and the magnetic field $h$ enter as a quadratic order term in Eq.~(\ref{spin cycloid solution}) with a negative sign and counteract the easy-axis anisotropy $\propto K_u$ and therefore can be interpreted as an effective easy-plane anisotropy. In particular, one special case occurs when the effective anisotropy vanishes, $K_{\text{eff}} = 0$, which yields:
\begin{align}
    \cos\theta = \cos\left(\frac{\alpha P_z}{2A}r\right) \, , \quad \sin\theta = \sin\left(\frac{\alpha P_z}{2A}r\right) \, .
\end{align}
This defines a critical magetic field $h_c \equiv \sqrt{2K_u / \chi - 4\beta^2}$, where the anharmonic effects vanish and therefore $\theta$ becomes linearly dependent on $r$ (see Eq.~(\ref{spin cycloid solution})).

Without loss of generality, let us assume $\phi = 0$, implying that the spin cycloid structure lies along the $x$-direction.
Then, $r$ can be simply replaced by $x$ and consequently, the antiferromagnetic vector $\mathbf{n}$ of the spin cycloid structure rotates within the $xz$-plane.
Moreover, if there is no external magnetic field the magnetization vector $\mathbf{m}$ aligns along the $y$-direction, exhibiting the identical periodicity as the spin cycloid structure.
The magnitude of $\mathbf{m}$ is fully determined by the cycloid structure.
In this case, it is evident that the $\mathbf{n}$ and $\mathbf{m}$ vectors exhibit no dependence on the $y$-direction.
The ground state of the system within the $xz$-plane is illustrated in Fig.~\ref{fig:Ground state}. Upon inserting the values from Table.~\ref{table:coefficient} for verification, it is observed that the value of $\mathbf{m}$ is approximately 20 times smaller than $\mathbf{n}$.
Hence, it can be inferred that there is a consistency with the assumed constraint, which establishes the antiferromagnetic nature of the material.

\subsection{Magnon excitation}

Form the Eq.~(\ref{Lagrangian}) and Eq.~(\ref{free energy 2}), the Euler-Lagrange equation for $\mathcal{L}$ can be derived as
\begin{align} 
    0 = &-sm_\theta\Dot{\phi}\cos\theta -s\partial_t m_\phi + 2A(\nabla^2\theta)  \nonumber \\
    &+2\alpha P_z \sin^2\theta\left[\sin\phi(\partial_x\phi)-\cos\phi(\partial_y\phi)\right]  \nonumber\\
    &- \sin2\theta [A(\nabla\phi)^2 + K_u] - 2\beta m_\phi \cos\theta \nonumber \\
    &- h m_\theta \cos\theta , \label{EOM1} \\
    0 = &s\partial_t(m_\theta \sin\theta) + 2A\nabla\cdot[\sin^2 \theta(\nabla\phi)] \nonumber \\
    &- 2\alpha P_z \sin^2 \theta[\sin\phi(\partial_x\theta) -\cos\phi(\partial_y\theta)] , \\ 
    0 = &-s\Dot{\phi} \sin\theta - \frac{m_\theta}{\chi} - h \sin \theta, \\ 
    0 = & s\Dot{\theta} - \frac{m_\phi}{\chi} - 2\beta \sin \theta , \label{EOM2}
\end{align}
in spherical coordinates. To obtain the spin-wave solution for small deviations from the ground state, we set
\begin{align}
    \theta &= \theta_g + \delta\theta \,\, , \quad \,\,\,\,\,\quad \quad \phi = \phi_g + \delta\phi , \, \nonumber \\
    m_\theta &= m_{\theta,g}(\theta) + \delta m_\theta \, , \, m_\phi = m_{\phi,g} (\theta) + \delta m_\phi , \nonumber
\end{align}
where the subscript $g$ refers to the configuration of the ground state, while $\delta \theta, \delta \phi, \delta m_\theta$, and $\delta m_\phi$ indicate the small deviations of the corresponding quantities from the ground state.

Next, we expand the above Eqs.~(\ref{EOM1}-\ref{EOM2}) up to linear order in $\delta\theta$, $\delta\phi$, $\delta m_\theta$, and $\delta m_\phi$.
To find the spin wave dynamics, we use the separation of variables method.
As described above, we set the ground state of the system to have the $x$-directional spin cycloid structure.
Thus, we take $\phi_g = 0$.
Rotating about the z-axis by $\pi$ radians allows us to represent the remaining cases ($\phi_g = \pi$) as well.
Applying the ansatz $\psi \sim f(x) e^{i(k_y y + k_z z -\omega t)}$, the equations are simplified.
In the case of uniform distribution along y-axis, such as in a thin film in the xz-palne, setting $k_y = 0$, we obtain the following expression :
\begin{align}
    0 =& [(s^2\omega^2 \chi - 2Ak_z^2) + 2A\partial_x^2 - 2 K_{\text{eff}} \cos2\theta_g]\delta\theta \, , \label{Can be analytically solve} \\
    0 =& [(s^2\omega^2 \chi - 2 A k_z^2) + 2 A \partial_x^2 +  2A (\partial_x \theta_g)^2 \nonumber \\
    & \qquad \qquad \,\,\, - 2\alpha P_z (\partial_x\theta_g) - 2 K_{\text{eff}} \cos^2 \theta_g ] \delta \eta \, , \label{May not be analytically solve} \\
    0 =& i s \omega \chi \delta\phi\sin\theta_g - \delta m_\theta \label{EOM5} \, , \\
    0 =& i s \omega \chi \delta \theta + \delta m_\phi \, , \label{EOM6}
\end{align}
where $\delta\eta = \sin\theta_g\delta\phi$. Then, Eq.~(\ref{Can be analytically solve}) and Eq.~(\ref{May not be analytically solve}) allow us to recognize two different modes of excitations.
Equation~(\ref{Can be analytically solve}) describes the in-plane oscillation of the $\mathbf{n}$ vector within the cycloid plane, referred to as the in-plane mode.
Equation~(\ref{May not be analytically solve}) depicts the out-of-plane oscillation of the $\mathbf{n}$ vector perpendicular to the cycloid plane, referred to as the out-of-plane mode.
A schematic picture of both modes can be found in Fig.~\ref{fig:Ground state}.
Additionally, knowing the in-plane and out-of-plane modes provides complete information about the dynamics of the magnetization vector, as described by Eq.~(\ref{EOM5}) and Eq.~(\ref{EOM6}).

\begin{figure}
    \centering
    \includegraphics[width=0.4\textwidth]{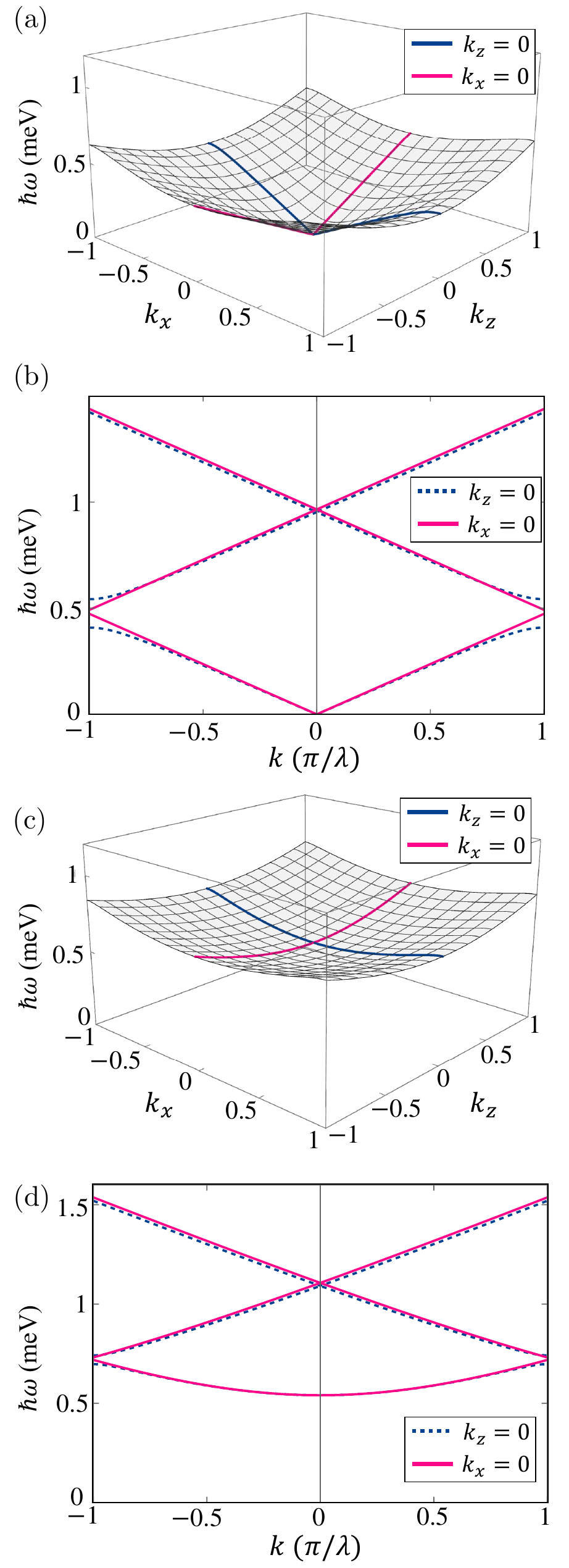}  
    \caption{The band structures of two different modes for zero external magnetic field: (a,b) the in-plane mode and (c,d) the out-of-plane mode.
    $\lambda$ is the wavelength of the periodic function, with a value of 61.05$\mathrm{nm}$.
    The blue dotted lines signify propagation along the $x$-direction ($k_z$ = 0), and the rose color solid lines depict propagation along the z-direction ($k_x$ = 0).
    Along the $z-$direction, the ground state spin configuration is uniform and therefore the plots in the figure show a trivial band folding along the z-direction
    }
    \label{fig:dispersion relation}
\end{figure}

Dispersion relation of in-plane mode can be obtained analytically by solving the Lamé equation (\ref{Lame}) for $n = 1$, $\nu = \sqrt{\frac{C}{A}}x$, $\kappa^2 = A K_{\text{eff}}/\alpha P_z^2$, $l = K_{\text{eff}}/C$ and $E = (s^2\omega^2 \chi - 2Ak_z^2)/2C$ :
\begin{align}
    \left\{\frac{d^2}{d\nu^2} - \kappa^2 [n(n+1) \text{sn}^2(\nu,l) -1 ] + E \right\} \delta\theta = 0 \, . \label{Lame}
\end{align}
The analytic solution of the excitation mode of Eq.~(\ref{Can be analytically solve}) is given by :
\begin{align}
    \delta\theta(\nu) = \frac{H(\nu \pm \gamma)}{\Theta(\nu)}e^{\mp \nu \mu Z(\gamma)} \, , \label{exactsol}
\end{align}
where $H(\nu)$, $\Theta(\nu)$, and $Z(\gamma)$ are Jacobi's eta, theta and zeta functions~\cite{Lame1, Lame2, KISHINE20151}.
In Eq.~(\ref{exactsol}), $\gamma = \text{dn}^{-1}(\sqrt{E},m)$ with the choice $\text{Im}{\gamma} > 0$ where dn is delta amplitude.
Since the potential is a periodic function, the solution must also be a periodic function by the Bloch theorem.
In this case, the period of the potential is half that of the spin cycloid.
Therefore, $Z(\gamma)$ is pure imaginary and the value for the $Z(\gamma)$ function can be determinded.

To obtain the dispersion relation of out-of-plane mode, we need to solve eigenvalue problem Eq.~(\ref{May not be analytically solve}).
In this case, utilizing the periodicity of the spin structure, it is possible to determine the excitation, referred to as the central equation~\cite{Ashcroft76, kittel2018introduction}.
\begin{align}
    \left[ \frac{d^2}{dx^2} + P(x)\right] f(x) = \epsilon f(x) \, ,
\end{align}
where $P(x)$, $\lambda$, and $\epsilon$, are respectively the periodic function, wavelength of the given periodic function and eigenvalue.
Here, the wavelength of the periodic function is half the wavelength of the spin cycloid.
Using the Fourier transformation,
\begin{align}
    P_n &\equiv \frac{1}{\lambda}\int_0^{\lambda}e^{-iqnx}P(x)dx , \\
    P(x) &= \sum_{n \in \mathbb Z} P_n e^{iqnx} , \\
    f(x) &= \sum_{k} a(k) e^{ikx} ,
\end{align}
where $q=2\pi/\lambda$.
We obtained the following equation.
\begin{align}
    \epsilon \sum_{k} a(k) e^{ikx} = & -\sum_{k} a(k) k^2 e^{ikx} \nonumber \\
    & \, + \sum_{m \in \mathbb Z} P_m e^{iqmx} \sum_{k} a(k) e^{ikx} .
\end{align}
Each Fourier component must have the same value on both sides, thus the central equation becomes
\begin{eqnarray}
    k^2 a(k) + \epsilon a(k) = \sum_{m\in\mathbb Z} P_m a(k-mq) .
\end{eqnarray}

Solving Eq.~(\ref{Can be analytically solve}) using the central equation method or solving the Lamé equation analytically, both yield the same dispersion relation.
However, we were unable to to find the closed-form solution of Eq.~(\ref{May not be analytically solve}), so we obtain its dispersion relation numerically by solving the central equation.
The values used in the numerical calculations are those for BFO~\cite{2014_park} given in Table.~\ref{table:coefficient}. Using the parameters in Table.~\ref{table:coefficient} in numerical computations, we generate Fig.~\ref{fig:dispersion relation}.
The wavelength $\lambda$ and $C$ are determined during the process of minimizing the average free energy density.

\begin{table}
    \begin{tabular}{c|c}
         Parameter& \qquad Value\\ \hhline{=|=}
         $K_u$& \qquad $6.71 \times 10^5$ erg $\cdot$ cm$^{-3}$ \\ \hline
         $\alpha P_z$& \qquad $1.37$ erg $\cdot$ cm$^{-2}$ \\ \hline
         $\beta$& \qquad $1.12 \times 10^7$ erg $\cdot$ cm$^{-3}$ \\ \hline
         A& \qquad $6.59 \times 10^{-7}$ erg $\cdot$ cm$^{-1}$ \\ \hline
         $\chi$&  \qquad $1.10 \times 10^{-9}$ erg$^{-1}$ $\cdot$ cm$^3$ \\ \hline
         $s$&  \qquad $4.98 \times 10^{-5}$ erg $\cdot$ s $\cdot$ cm$^{-3}$
    \end{tabular}
    \caption{Values of material constants used in numerical calculation~\cite{2014_park}.}
    \label{table:coefficient}
\end{table}

In Fig.~\ref{fig:dispersion relation}, we observe the anisotropy of the bands.
The magnon band in the $x$-direction (with $k_z = 0$) exhibits a gap $\Delta$ at the Brillouin-zone boundary which arises due to the periodcity of the spin cycloid, while the magnon band in the $z$-direction (with $k_x = 0$) does not have a gap. Note that the gap $\Delta$ exists only between the lowest band and the second lowest band in the $x$-direction band (with $k_z = 0$) for both modes, i.e. at higher bands, no gaps are observed at the boundary or center of the Brillouin zone in the $x$-direction. The magnon band in the $z$-direction in-plane mode follows a normal antiferromagnetic dispersion relation.

Another noteworthy point is that the in-plane mode represents a massless excitation (Fig.~\ref{fig:dispersion relation}(b)).
The gapless at the zone center is the Goldstone mode associated with the spontaneous symmetry breaking.
In the ground state depicted in Fig.~\ref{fig:Ground state}, translational symmetry in the $x$-direction is broken by the magnetic ground state, while it remains preserved in the Hamiltonian state. Hence, in this model, spontaneous symmetry breaking occurs, and the in-plane mode of magnons corresponds to the Goldstone boson.

\begin{figure}
    \centering
    \includegraphics[width=0.48\textwidth]{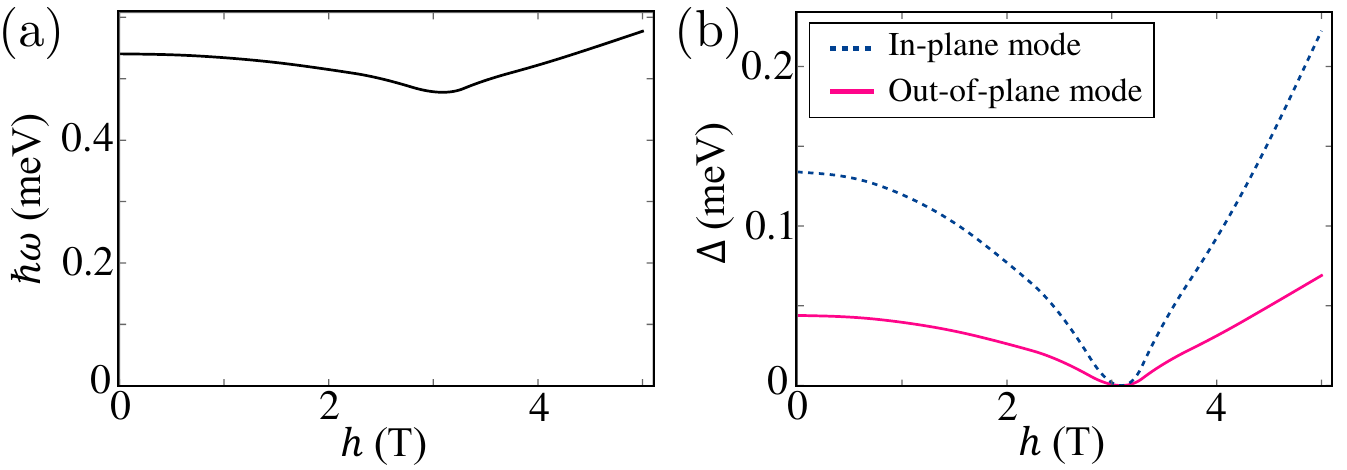}
    \caption{(a) Band gap $\hbar \omega$ of the out-of-plane mode at the center of the Brillouin zone as a function of the magnetic field $h$. (b) Band gap $\Delta$ between the lowest and next-lowest bands at the Brillouin zone edge in the $x$-direction for each mode as a function of the magnetic field $h$. The red line represents the out-of-plane mode, and the blue dashed line represents the in-plane mode.}
    \label{fig:Gap}
\end{figure}

Figure~\ref{fig:Gap} illustrates how the band gap changes with the magnetic field. At the center of the Brillouin zone, the in-plane mode always has zero energy, while the out-of-plane mode reaches its minimum value at the critical field, determined by $h_c = \sqrt{(2K_u - 4\beta^2\chi)/\chi}$, which is about $3.07 \, \mathrm{T}$  for BFO, and increases on either side of this field as shown in Fig.~\ref{fig:Gap}(a). At the critical field, the gap $\hbar \omega$ at the Brillouin-zone center has the value $\hbar \alpha P_z / (\sqrt{2\chi A} s)$, which is about 0.5 $\mathrm{meV}$ for BFO. Figure~\ref{fig:Gap}(b) shows the gap $\Delta$ between the lowest and the second lowest band at the boundary of the Brillouin zone along x, where the gap closes at the critical field but opens and increases as one moves away from the critical field.
This implies that the anisotropy inherent in the material can be controlled via the magnetic field.
The field dependence of the gap $\Delta$ as a function of a magnetic field is one of our key findings of this work.

\subsection{Magnon thermal conductivity}

We utilize the Boltzmann transport equation within the relaxation time approximation to investigate thermal conductivity of thermal magnon.
The equation takes the form
\begin{align}
    \frac{\partial g}{\partial t} + \mathbf{v}\cdot\nabla g + \dot{\mathbf{k}} \cdot \nabla_k g = - \frac{g - g_0}{\tau} .
\end{align}
Here, $\tau$ denotes the relaxation time, and $g$ and $g_0$ represent the magnon distribution functions of the non-equilibrium and equilibrium states, respectively, where $g_0(\epsilon,T) = (e^{\epsilon/k_BT}-1)^{-1}$. Focusing on the stationary $g$, i.e., $\partial_t g = 0$.
The Boltzmann equation is simplified to
\begin{equation}
     \mathbf{v}\cdot\nabla g = - \frac{g - g_0}{\tau} .
\end{equation}

Assuming that the thermal gradient of the given sample is sufficiently small, the non-equilibrium distribution function can be understood as a small variation from the equilibrium distribution function and can be expressed as
\begin{equation}
    g(\epsilon,T,\nabla T) = g_0(\epsilon,T) + g_1(\epsilon,T,\nabla T) .
\end{equation}
Then, after combining it with Boltzmann transport equation, we can obtain
\begin{equation}
    g_1 = - \tau \mathbf{v}\cdot\nabla g_0 - \tau \mathbf{v}\cdot\nabla g_1.
\end{equation}
Under the assumption that the thermal gradient of the given sample is sufficiently small, $\nabla g_1$ can be ignored to compare $\nabla g_0$, leading to the following result.
\begin{equation}
    g_1 \approx - \tau \mathbf{v}\cdot\nabla g_0 = - \mathbf{v}\cdot\nabla T \tau \frac{\partial g_0}{\partial T} .
\end{equation}
The heat flux flowing in the $\alpha$ direction is
\begin{equation}
    j^{heat}_\alpha = \int d\mathbf{k} \epsilon(\mathbf{k})v_\alpha(\mathbf{k})g(\epsilon,T,\nabla T).
\end{equation}
And we know that $ j^{heat}_\alpha = -\kappa_{\alpha\beta} (\partial_\beta T)$, where
\begin{equation}
    \kappa_{\alpha\beta} = \int d\mathbf{k} \epsilon(k)v_\alpha(k)v_\beta(k) \tau \frac{\partial g_0}{\partial T}.
\end{equation}
In numerical calculations, the magnon velocity is defined as $v_\alpha(k) = \frac{1}{\hbar}\frac{d \epsilon}{d k_\alpha}$ and the relaxation time is assumed to be constant at $\tau = 10^4 \, \mathrm{ns}$.

\begin{figure}[t]
    \centering
    \includegraphics[width=\columnwidth]{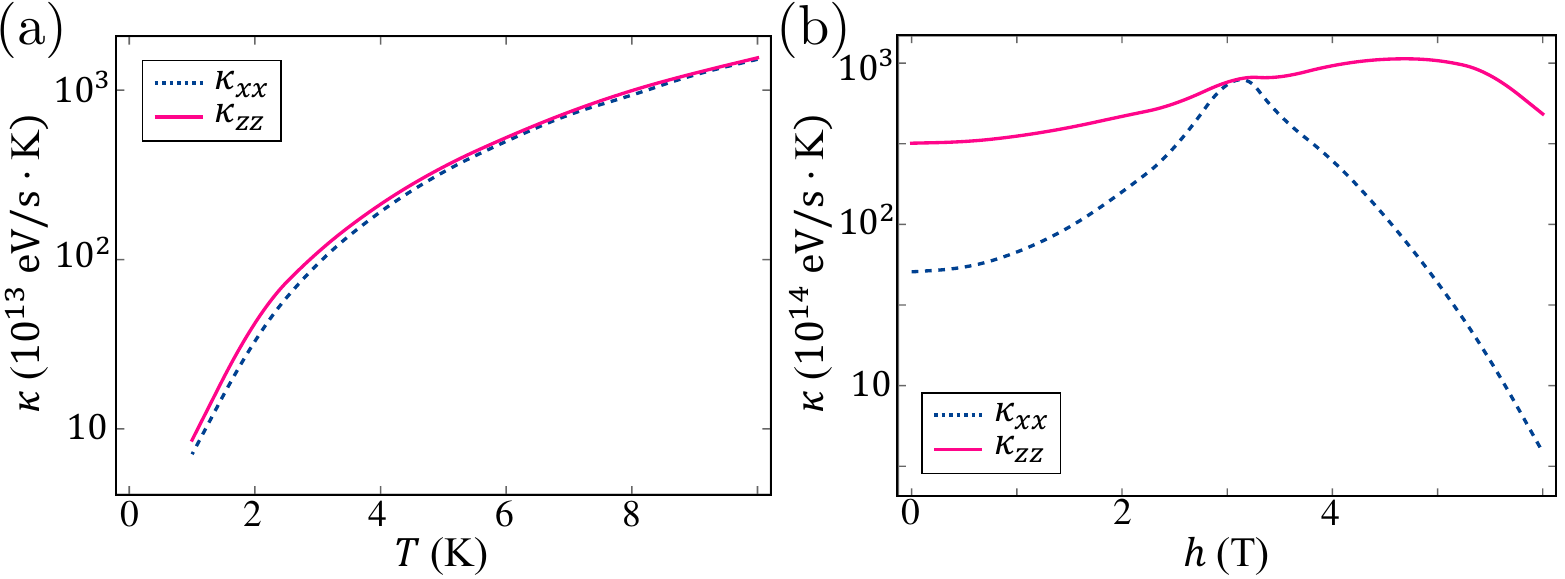}
    \caption{(a) Thermal conductivity of a thin film BFO at a zero magnetic field as a function of temperature $T$. (b) Thermal conductivity of a thin film BFO at 3K as a function of a magnetic field $h$. The dashed blue line and the solid red line represent $\kappa_{xx}$ and $\kappa_{zz}$, respectively.}
    \label{fig:Thermal conductivity}
\end{figure}

In Fig.~\ref{fig:Thermal conductivity}(a), the numerical result demonstrates anisotropy of thermal conductivity between the $x$- and $y$-directions.
The anisotropy of thermal conductivity is most noticeable at low temperatures because anisotropy of band is most pronounced in the lowest band, becoming less prominent in higher bands as the band gap and anisotropy become negligible relative to the temperature scale.
In other words, as we move towards higher temperatures, the contribution of bands with reduced anisotropy to thermal conductivity increases, leading to diminished anisotropy of thermal conductivity. In Fig.~\ref{fig:Thermal conductivity}(b), we can see that the thermal conductivity changes as a function of the magnetic field.
This shows that the anisotropy, governed by the magnetic field, also affects transport properties.
In particular, $\kappa_{xx}$ shows the peak around the critical field $h= \sqrt{(2K_u - 4\beta^2\chi)/\chi} \sim 3.07 \, \mathrm{T}$, which stems from the field dependence of the magnon bands shown in Fig.~\ref{fig:Gap}(b) and therefore can be used to infer the underlying spin-cycloid structures.

\section{Conclusion} \label{section4}
We have investigated the spin-cycloid ground states and the spin-wave excitations of multiferroic antiferromagnets, taking BFO as a paradigmatic example of an antiferromagnetic material in which an inversion-symmetry-breaking distortion can induce a spin cycloid. Our analysis has revealed the two distinct modes of excitations within the band structure, including one massless excitation, attributed to the Goldstone boson resulting from spontaneous translational symmetry breaking of the ground-state spin cycloid that exhibits gap opening at the magnetic Brillouin zone boundary. We have found that the effect of the DMI and the magnetic field (aligned with the polarization vector) on the dynamics of magnons can be interpreted as the effect of the effective easy-plane anisotropy. To study the transport properties, the Boltzmann transport equation under the relaxation time approximation is used to investigate thermal conductivity. Our numerical results have shown the anisotropy in thermal conductivity, particularly at low temperatures, which is attributed to the pronounced anisotropy in the band structure.
However, at higher temperatures, the contribution of bands with reduced anisotropy becomes more significant, resulting in diminished overall anisotropy in thermal conductivity. These findings highlight the temperature-dependent anisotropic behavior of thermal conductivity. We have also investigated the magnetic field dependence of the magnon bands and the thermal conductivity. These findings contribute to the understanding of multiferroic materials, shedding light on the dynamics and behaviors of magnons within these complex systems.

\begin{acknowledgments} H.W.P. and S.K.K. were supported by Brain Pool Plus Program through the National Research Foundation of Korea funded by the Ministry of Science and ICT (NRF-2020H1D3A2A03099291), by the National Research Foundation of Korea (NRF) grant funded by the Korea government (MSIT) (NRF-2021R1C1C1006273), and by the National Research Foundation of Korea funded by the Korea Government via the SRC Center for Quantum Coherence in Condensed Matter (NRF-RS-2023-00207732). S. Z. is supported by the Leibniz Association through the Leibniz Competition Project No. J200/2024. S.H. and R.R. acknowledge funding from the U.S. Department of Energy, Office of Basic Energy Sciences, Materials Sciences and Engineering Division under Contract No. DE-AC02-05-CH11231 (Codesign of Ultra-Low-Voltage Beyond CMOS Microelectronics) for the development of materials for low-power microelectronics. M.R. S.H., and R.R. acknowledge Air Force Office of Scientific Research 2D Materials and Devices Research program through Clarkson Aerospace Corp under Grant No. FA9550-21-1-0460. S.H. and R.R. also acknowledge the ARO-CHARM program as well as the NSF-FUSE program. We also acknowledge the Army Research Laboratory and was accomplished under Cooperative Agreement Number W911NF-24-2-0100. The views and conclusions contained in this document are those of the authors and should not be interpreted as representing the official policies, either expressed or implied, of the Army Research Laboratory or the U.S. Government. The U.S. Government is authorized to reproduce and distribute reprints for Government purposes notwithstanding any copyright notation herein. J.Í.-G. acknowledges the ﬁnancial support from the Luxembourg National Research Fund through grant C21/MS/15799044/FERRODYNAMICS.
\end{acknowledgments}

\bibliography{ref}

\end{document}